\documentclass[useAMS,usenatbib]{mn2e}
\usepackage{natbib}
\usepackage{graphicx} 
\usepackage[usenames,dvipsnames,svgnames,table]{xcolor}
\usepackage{ulem}
\usepackage{amsmath}

\newcommand{\aaps}[1]{A\&AS}
\newcommand{\solphys}[1]{Solar Physics}
\newcommand{\aapr}[1]{Aapr}
\newcommand{\apj}[1]{ApJ}
\newcommand{\mnras}[1]{MNRAS}
\newcommand{\prd}[1]{prd}

\title[BiSON data preparation]{BiSON data preparation: A correction for differential extinction and the weighted averaging of contemporaneous data}
\author[G. R. Davies, W. J. Chaplin, Y. Elsworth, and S. J. Hale]{G. R. Davies $^{1}$\thanks{E-mail:
davies@bison.ph.bham.ac.uk}, W. J. Chaplin$^{1}$, Y. Elsworth $^{1}$, and S. J. Hale $^{1}$.\\
$^{1}$HiROS/BiSON, University of Birmingham, Edgbaston, Birmingham, B15 2TT, UK }

\begin{document}

\date{Accepted 1988 December 15. Received 1988 December 14; in original form 1988 October 11}

\pagerange{\pageref{firstpage}--\pageref{lastpage}} \pubyear{2013}

\maketitle

\label{firstpage}
\begin{abstract}
The Birmingham Solar Oscillations Network (BiSON) has provided high-quality high-cadence observations from as far back in time as 1978.  These data  must be calibrated from the raw observations into radial velocity and the quality of the calibration has a large impact on the signal-to-noise ratio of the final time series.  The aim of this work is to maximise the potential science that can be performed with the BiSON data set by optimising the calibration procedure.  To achieve better levels of signal-to-noise ratio we perform two key steps in the calibration process: we attempt a correction for terrestrial atmospheric differential extinction; and the resulting improvement in the calibration allows us to perform weighted averaging of contemporaneous data from different BiSON stations.  The improvements listed produce significant improvement in the signal-to-noise ratio of the BiSON frequency-power spectrum across all frequency ranges.  The reduction of noise in the power spectrum will allow future work to provide greater constraint on changes in the oscillation spectrum with solar activity.  In addition, the analysis of the low-frequency region suggests we have achieved a noise level that may allow us to improve estimates of the upper limit of g-mode amplitudes.
\end{abstract}


\section{Introduction}
The 35 year old BiSON \citep{1996MNRAS.282L..15C} is an observational network, unique in its temporal baseline, that allows us to peel back the outer layers of the Sun and observe the interior solar structure and dynamics.  BiSON's extended observations allow us to resolve long oscillation mode lifetimes while the historic nature of observations provides insight on the Sun during three solar activity cycles.  The importance of the data set should not be underestimated - a new Sun-as-a-star observatory would require time scales of an order of an academic lifetime to reach a similar maturity.  For this reason optimising the existing data set offers ``value for money''.  This paper details the recent work applied to producing the best possible BiSON data set.\\
BiSON is a six-station ground-based network attempting to continuously monitor the Sun.  Local station weather and mechanical failure play a role in the ability of each station to produce data.  At any given epoch, observations may be taken by between zero and four stations.  It is the goal of the network to produce a single time-domain data set that has a high duty cycle, thus minimising the impact of the temporal window function \citep{1993A&A...280..704L}, while maintaining a high signal-to-noise ratio.\\  
The impact of the window function is well known: a reduction in signal-to-noise ratio, increased correlation of previously near independent frequency bins, and a modification of the underlying estimated limit spectrum (for details see \cite{2011arXiv1103.5352A}).  Especially for a ground based network, periodic interruptions cause diurnal side bands in the frequency domain.  Here we propose a correction to the raw observations that allows BiSON to record an increased duty cycle though improvements to the calibration of data at the extreme ends of the day.  Hence, this proposed correction helps to mitigate the impact of the window function.\\
More than simply improving the duty cycle, this paper demonstrates that a significant step has been taken in the merging of contemporaneous BiSON velocity residuals.  Dating back to the late 1990's it has been the BiSON practice to concatenate time series from individual sites by selecting only the data from the ``best'' station available \citep{1997A&AS..125..195C}.  Until now, this methodology was the only available if one wanted to study the low-frequency regime.  Diurnal drifts caused by end-of-day terrestrial atmospheric extinction prevented the averaging of multi-site data.  However, with a correction for these drifts caused by atmospheric extinction we may reintroduce the weighted averaging of multi-site overlaps.\\
This paper provides a method for determining the impact of terrestrial atmospheric noise and demonstrates a correction for some of these effects.  Here we make a distinction between forms of atmospheric noise: random noise due to atmospheric disturbances which requires real time monitoring for correction \citep{1988ESASP.286..223M,1994ExA.....4..253K}; systematic atmospheric refraction that is differential across the solar disk and considered to be a small effect for all but the very extreme ends of the day; and systematic noise due to the atmospheric gradient across the solar disc, monotonically changing as a function of solar zenith angle.  This last systematic effect is commonly referred to as `differential extinction', differential in space across the solar disc, and is the effect considered here.\\
The motivation for a better signal-to-noise ratio is the study of the seismic solar cycle and the detection of low-frequency modes to give better constraints on solar structure and dynamics.  A better signal to noise would allow us to analyse shorter data sets giving better temporal resolution to study how the Sun changes with the activity cycle.  In the low-frequency region, both pressure and gravity modes (p and g modes) may be detectable and measurable with exquisite accuracy and precision.  The unambiguous detection of individual g modes, with very small uncertainties, would provide constraints on the solar core \citep{2009ApJ...699.1403B}, dark matter \citep{2012ApJ...746L..12T, 2010PhRvD..82j3503C,2010ApJ...722L..95L}, and the stochastic gravitational wave background \citep{2014arXiv1401.6888S}.\\  
The remainder of the paper is laid out as follows.  Section \ref{methods} describes the correction for differential extinction and the weighted averaging of contemporaneous data, and details current efforts in gap filling.  Section \ref{results} shows the impact of the work applied to real data and quantifies the improvements in signal-to-noise ratio.  Section \ref{conclusions} draws conclusions on the efficacy of the work and discusses the science that may result.  The Appendix details the numerical model used to study the impact of differential extinction on the BiSON resonance scattering spectrometer data. 
\section{Methods}
\label{methods}
\subsection{Terrestrial atmospheric differential extinction}
BiSON measures the solar Doppler velocity field by taking unresolved (Sun-as-a-star) observations of the solar-surface radial-velocity field using resonance scattering spectrometers (RSS) \citep{1978MNRAS.185....1B}.
This measurement is achieved by referencing the Doppler shifted solar potassium D1 Fraunh\"ofer line, at around 770 nm, to the same atomic transition in a laboratory frame of reference.  By applying a magnetic field of approximately $0.18 \rm \; T$ to the resonance scattering medium, the laboratory frame transition is Zeeman split by $5200 \rm \; m \; s^{-1}$. Hence the transition between the Zeeman shifted $4^{2}S_{\frac{1}{2}}$ and $4^{2}P_{\frac{1}{2}}$ states produces four Zeeman components of which two are measured, the blue shifted $\sigma_{-}$ line ($\Delta m_j = -1$), referred to as the blue component, and the $\sigma_{+}$ line ($\Delta m_j = +1$), known as the red component.  The instrument switches between the red and blue component by varying the state of an electro-optic analyser.  Switching at $50 \rm \, Hz$ helps to overcome the changing state of the terrestrial atmosphere.  Critically for good instrumental stability, the difference in the blue and red intensity measurements divided by the their sum defines a ratio that is a good measure of the Sun-observer line-of-sight velocity \citep{1995A&AS..113..379E}.\\
This observational technique takes raw measurements that are intensities (i.e. a photometric measure) and calibrates these to produce Doppler velocities.  But of course, it is well known that ground-based photometry is subject to noise produced by terrestrial atmospheric extinction, and BiSON instruments are no exception.\\
Atmospheric extinction is a significant source of noise in the low-frequency regime and its removal would assist in the detection of low-frequency solar oscillations.
However, these modes suffer from low signal to noise in the frequency domain, making detection a significant challenge \citep{2010A&ARv..18..197A, davies2014}. \\

Differential extinction is an artefact of
the gradient of atmospheric extinction across the solar disc, the disc being
a discrete source.  The magnitude and alignment of this gradient causes
changes in measured intensities that are different for the red and
blue components.  In other words, because the solar disc is rotating, the whole-disc solar absorption line is rotationally broadened.  By placing a systematic transmission gradient at some angle to the rotational axis, the whole disc
solar line that is observed is given an asymmetry that manifests in the measured intensities.  If the weighting is more heavily biased to the receding limb, the red wing of the solar line is enhanced as the blue wing fades, and vice versa.\\
The systematic signal found in the measured intensities persists through the calibration process that reduces raw data to velocities.  In the worst case scenario, the differential atmospheric signal may be confused with signal of solar origin \citep{1979A&A....77..351G,1980A&A....88..317S}.  In addition, the atmospheric signal places significant low-frequency noise into the power spectrum reducing the already low signal-to-noise ratios of modes in that part of the frequency regime. \citep{1988ESASP.286..177B,1991SoPh..133...81E}.  It is clear that a robust correction for differential extinction is desirable.\\

\subsubsection{The correction through calibration}
The BiSON calibration pipeline takes daily ratios $R$, that include source-observer line-of-sight velocity components $v_{\rm{station}}$, and fits a third-order polynomial in ratio-velocity space, with coefficients $a_{i}$ where $0 \leq i \leq 3$ \citep{1995A&AS..113..379E} i.e.
\begin{equation}
R(t) = \sum_{i=0}^{3} a_{i} v(t)^{i}_{\rm{station}}.
\end{equation}
To convert from a ratio to a residual velocity $v_{\odot}$, the fit is subtracted from the measured ratio and multiplied by the inverse sensitivity $\rm{d}\it{v}_{\rm{station}}/\rm{d} \it{R}$.  The sensitivity is the rate of change of the ratio with respect to the station velocity,
\begin{equation}
\frac{{\rm d} R}{{\rm d} v_{\rm station}} = \sum_{i=1}^{3} i a_{i} v^{i-1}_{\rm{station}},
\end{equation}
giving the calibrated residual velocity as,
\begin{equation}
v_{\odot}(t) = \frac{R_{\rm{measured}} \it - \sum_{i=0}^{3} a_{i}v^{i}_{\rm{station}}}{\sum_{i=1}^{3} i a_{i} v^{i-1}_{\rm{station}}}.
\end{equation} 
In a previous method, the third-order polynomial fit attempted to minimise the impact of differential extinction by adjusting the fit to minimise the $\chi^{2}$ goodness-of-fit parameter.  The fit could not adjust to the shape of the differential extinction and so the differential extinction passed through the fit, albeit with some modification.  For this reason any correction for differential extinction must be applied before or at the same time as the fit.\\  
It is possible to determine the magnitude of the differential extinction using information from the model detailed in the Appendix or by calculating the extinction coefficient using Bouguer's method (see \cite{1961SvA.....5..399K}).  The Appendix provides detail and context on the instrument, measurements, and differential extinction.  However, rather than simulating from first principles a correction, it is also possible to modify the calibration procedure to include a differential extinction term that can be fitted simultaneously to the third-order polynomial.  We can determine the shape of the correction and use the fit to determine the correction magnitude.  This method has the benefit of fitting the correction to the data, but with a well defined shape that will minimise the removal of signal of solar origin.\\  
The shape of the differential extinction component has been studied previously (see
\cite{1979A&A....77..351G,1988ESASP.286..177B,1988ESASP.286..223M}).
It is common to seek a model for differential extinction that takes the
magnitude of the atmospheric transmission gradient
$T^{\prime} / T$ ($T$ is the transmission and $T^{\prime}$ is the first derivative) and multiplies this by the $\rm{sine}$ of the
angle between  
this gradient and the rotation axis of the Sun.  Including some
constant, $A_{\rm{inst, v}}$, to account for the instrumental sensitivity gives a
generic model for the velocity component of differential extinction I.e.,
\begin{equation}
v_{\rm{dext}} = A_{\rm{inst, v}} \frac{T^{\prime}}{T} \sin(q \pm P_{a}),
\end{equation}
where $q$ is the parallactic angle and $P_{a}$ is the position angle of the Sun.
The ambiguity in the plus or minus of these angles comes from an
ambiguity in definition.  Here we define the position angle such that
the angle between the transmission gradient and the solar rotation axis
is $q - P_{a}$.\\
\cite{1991SoPh..133...81E} realised that the empirical
factor $A$ would not remain constant through the year and so modelled
this parameter in detail.  The result of this work was a
complicated function containing many constants that attempted to
describe differential extinction throughout the year.  By including our correction in the calibration procedure we require only a function that remains valid throughout each day. \\
Here we make a subtle but necessary change to the approach of
previous authors, to derive a correction in terms of ratio not
velocity.  Previous work had adjusted the calibration procedure so
that $v_{\rm station}$ included the pseudo velocity from differential
extinction.  We derive an additional function that may be
fitted
to data at the same time as a third-order polynomial.  This allows for
the possibility of free parameters in the fit to account for
empirical instrumental factors and the poorly known atmospheric extinction
coefficient.\\ 
Given that we consider the existing model in velocity to be sensible,
we can straightforwardly transmute previous work to suit our needs by describing differential extinction in terms of the measured ratio:
\begin{equation}
r_{\rm{dext}} = A_{\rm{inst, r}} \frac{T^{\prime}}{T} \sin(q - P_{a}),
\end{equation}
which requires no more than describing the instrumental sensitivity, $A_{\rm inst, r}$ in terms of the ratio rather than velocity.\\
We use a model for the atmospheric transmission function that can be
expressed in terms of zenith angle, quoted directly from \cite{1989ApOpt..28.4735K},
\begin{equation}
T = \exp\left(-\kappa \frac{1}{\cos(z_{0}) -
0.50572\left(96.07995 - z_{0}\right)^{-1.6364}}\right),
\end{equation}
where the zenith angle $z_{0}$ must be given in degrees.  
For convenience we will express the
magnitude of the atmospheric transmission gradient with respect to
$z_{0}$ as the multiplication of just two terms,
\begin{equation}
\frac{T^{\prime}}{T} = -\kappa \beta(z_{0}),
\end{equation}
where
\begin{equation}
\beta(z_{0}) = 
\left(\frac{0.82756}{w^{2.6364}}-\sin(z_{0})\right)\left(\frac{0.50572}{w^{1.6364}}+\cos(z_{0})
\right)^{-2},
\end{equation}
and $w = 96.07995 - z_{0}$.  
This then gives our model in terms of ratio as
\begin{equation}
r_{\rm{dext}} = -A_{\rm{inst, r}} \kappa \beta(z_{0}) \sin\left(q - P_{a}\right).
\label{model1111}
\end{equation}
It is troublesome to measure the atmospheric extinction coefficient $\kappa$, especially as values may be quite different for morning and afternoon, but we require this value for Equation \ref{model1111}.  In addition, we require the
empirical instrumental sensitivity factor, $A_{arm inst, r}$.  It is therefore sensible to combine two
separate unknowns to form a single unknown, here $\alpha$.  We then
have a correction that is a function of zenith angle, parallactic
angle, and position angle with a single free parameter $\alpha$, I.e.,
\begin{equation}
r_{\rm{dext}}(z_{0},q,P_{a}) = \alpha \beta(z_{0}) \sin\left(q - P_{a}\right).
\label{model2222}
\end{equation}
This provides us with a function that can be combined with a
third-order polynomial to describe the measured ratio thus,
\begin{equation}
R(v,z_{0},q,P_{a}) = \sum_{i=0}^{3}a_{i}v^{i} + \alpha \beta(z_{0}) \sin\left(q - P_{a}\right).
\end{equation}
The data can then be calibrated by subtracting the fitted function and multiplying by the inverse sensitivity.\\
When applying this correction to real data, we find significant improvement in the correction for differential extinction when we assign two values of $\alpha$, one for morning and one for afternoon.  This allows for the varying atmospheric conditions through the day.  This simplistic approach works because the differential extinction correction is sensitive to the atmospheric extinction coefficient at the extremes of the day, but is insensitive at the middle of the day.  Hence, although the changing atmospheric conditions are best described by some function, the response of the correction to the same conditions can be well approximated by just two values.\\ 
\subsection{The merging of contemporaneous multi-site velocity residuals}
Residual velocities in station overlaps can be combined using a statistically weighted mean value.  This requires us to define a parameter to assess the required weight, W.  A focus of this work is the reduction in noise levels in the frequency region from $800$ to $1300 \rm \, \mu Hz$ in order to detect low-frequency p modes.  In this region the total signal is much less than the total noise, and so we can then define the noise parameter as the mean of the amplitude squared of the frequency-power spectrum in the defined region, $\sigma^2$.  The weight will simply be the inverse of the noise parameter, $W \approx 1 / \sigma^2$.\\
To combine the residual velocities, let the weight for a station in an overlap be $W_{j}$, and let there be $n$ stations such that $1 \geq j \geq n$.  The residual velocity from each station, $v_{j}(t)$, can be combined in the time domain to give the combined residual velocity $v_{\rm com}(t)$, according to    
\begin{equation}
v_{\rm com}(t) = \frac{\sum_{j=1}^{n} v_{j}(t) W_{j}}{\sum_{j=1}^{n}W_{j}}.
\end{equation}
\cite{1997A&AS..125..195C} showed that if each station in an overlap has an incoherent white noise source characterised by $\sigma_{j}^{2}$, and a signal where the height of the Lorentzian $H_{\nu}$ describing a mode is common to each station, then the signal-to-noise ratio in the power spectrum, $S/N$, of the mode for the duration of the overlap is,
\begin{equation}
S/N \approx \frac{H_{\nu} N}{2} \sum_{j=1}^{n}\frac{1}{\sigma_{j}^2},
\end{equation}
where $N$ is the number of points in the time series.  Hence, because the a weighted combination of overlap residuals from two similar sites will lead to an improvement in signal-to-noise ratio of a factor of two.\\
In reality, this is tempered by the fact that not all BiSON stations are of the same quality and that there are varying amounts of overlap.  However, it is clear that the weighted averaging of contemporaneous multi-site data is desirable if we have dealt with differential extinction to such an extent that diurnal drifts in the residual velocities have largely been removed.\\  

\section{Results}
\label{results}
The results presented here validate the methods used and demonstrate the improvement in the noise characteristics that will enable better observations at lower frequencies.  
\subsection{Single station daily impact}
\begin{figure}
\resizebox{\hsize}{!}{
\includegraphics{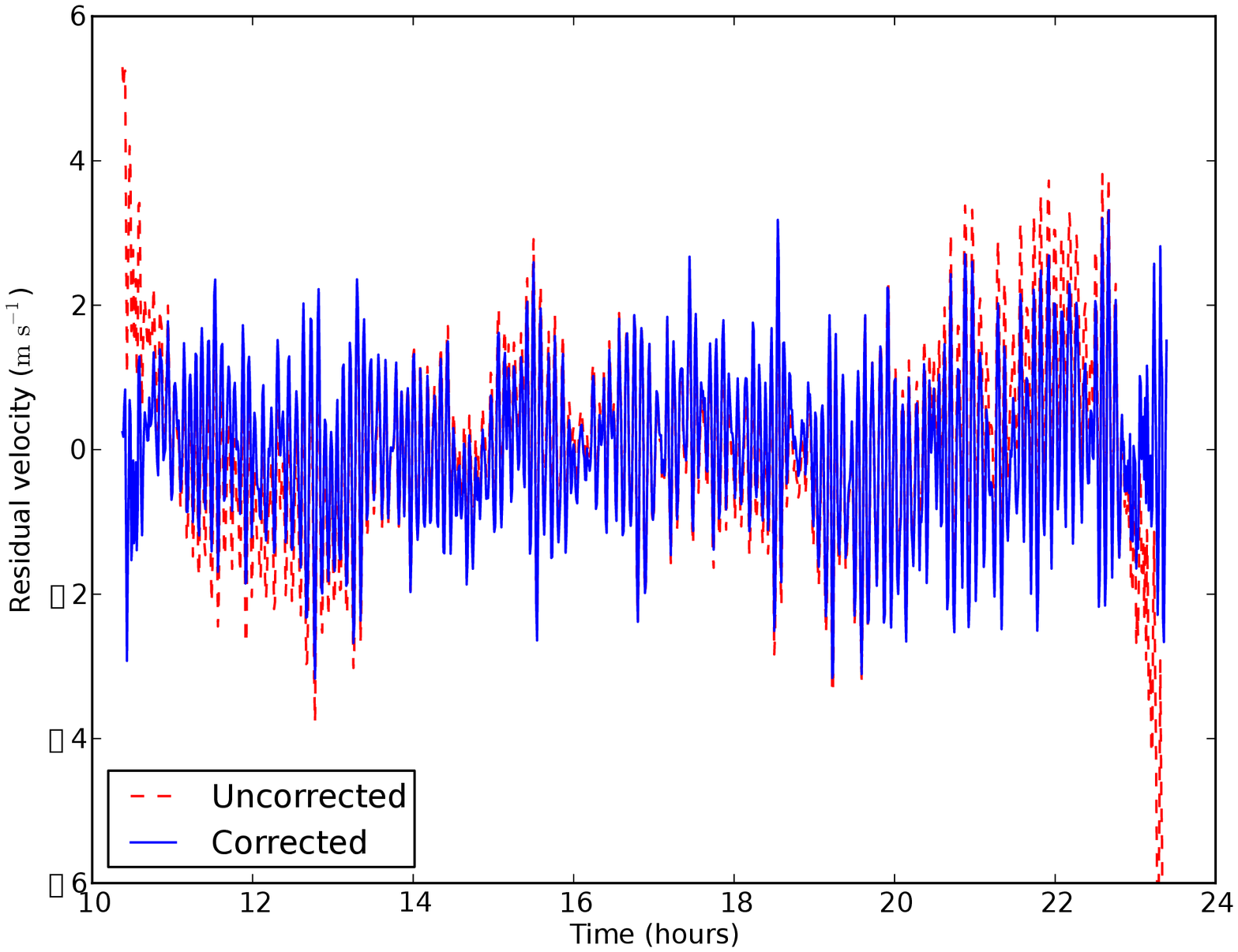}}
\resizebox{\hsize}{!}{
\includegraphics{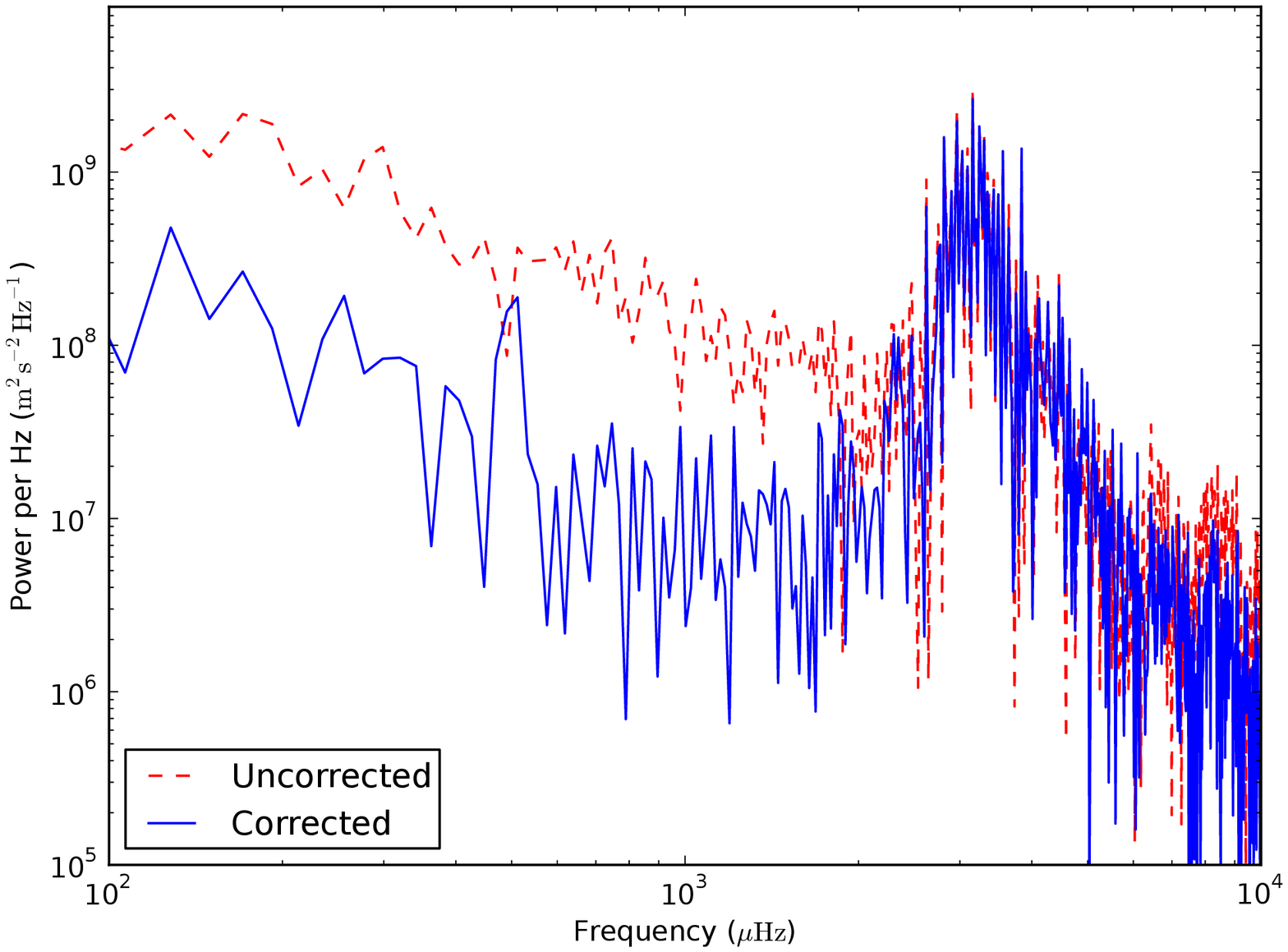}}
\caption{Calibrated data from the BiSON Las Campanas station December 28 2007 plotted with and without the correction for differential extinction.  Upper panel shows the time series and the lower panel the power spectrum.}
\label{fig::single_day}
\end{figure}

Figure \ref{fig::single_day} shows the calibrated residual velocities (in both the time and frequency domain) both with and without the correction for differential extinction.  BiSON data would normally be rejected at the extreme ends of the day in the absence of a differential extinction correction, hence comparing the wild systemic departures at the ends of the day with corrected data overstates the improvement.  However, it is our purpose to demonstrate that the correction for differential extinction is valid and this is most aptly done using data from the extreme ends of the day.\\
The improvement seen in the residual velocities is obvious.  In the time domain the residuals are ``flatter'' and this propagates through to the frequency domain as an order-of-magnitude reduction in low-frequency noise.  But more than this, the noise levels across the frequency range are improved.  This reduction in noise level, albeit exaggerated here, will improve the signal-to-noise ratio and give better access to the difficult to detect signal, such as the low-frequency p modes \citep{davies2014}, very-low-frequency g modes \citep{2007Sci...316.1591G} or the high-frequency pseudo modes \citep{2003ESASP.517..247C}.\\
\subsection{Common signal from different stations}
\begin{figure*}
	\includegraphics[width=177mm]{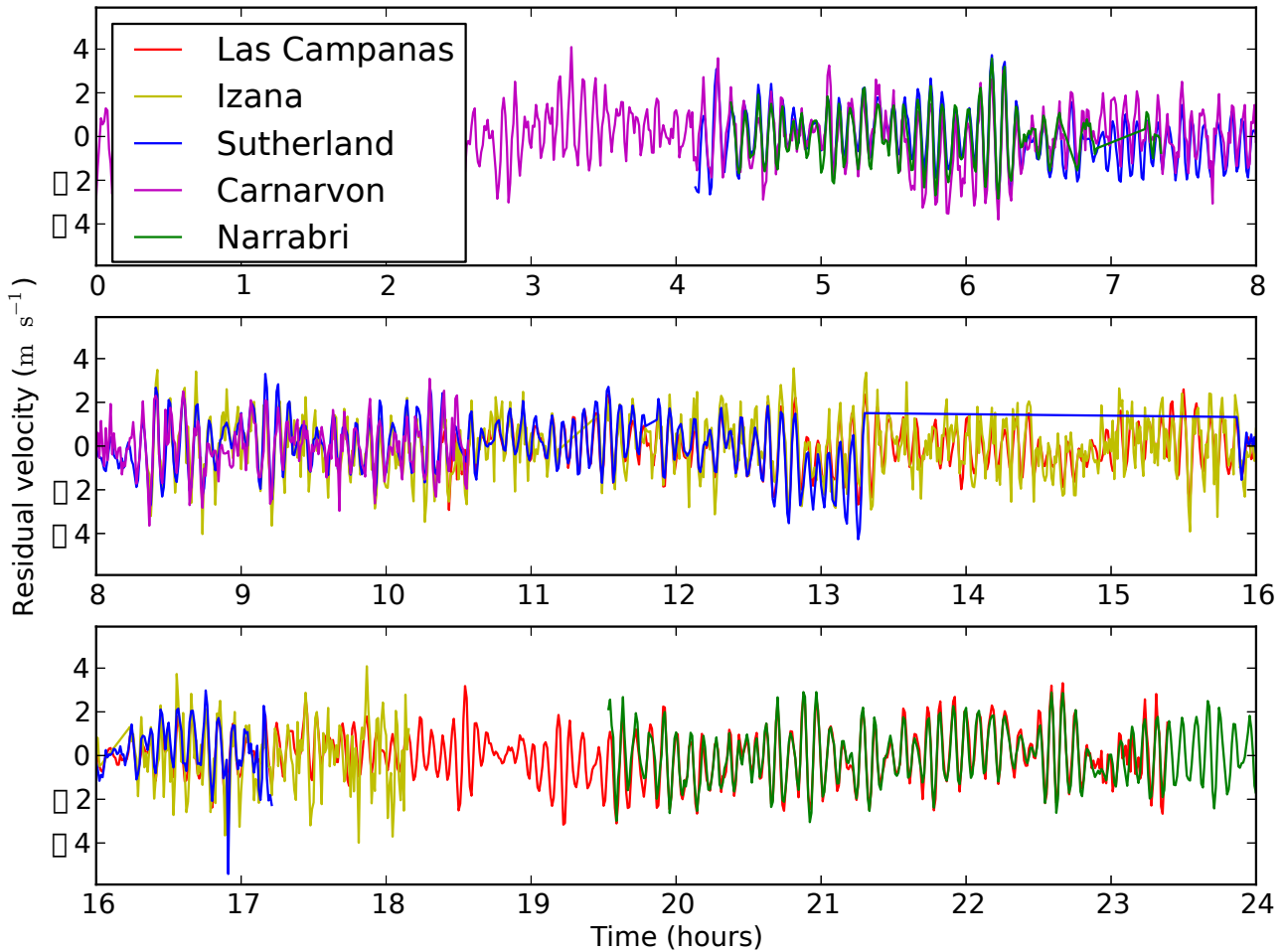}
	\caption{Data from the BiSON network on December 28 2007 calibrated with the correction for differential extinction.}
\label{fig::golden}
\end{figure*}

\begin{figure}
	\includegraphics[width=84mm]{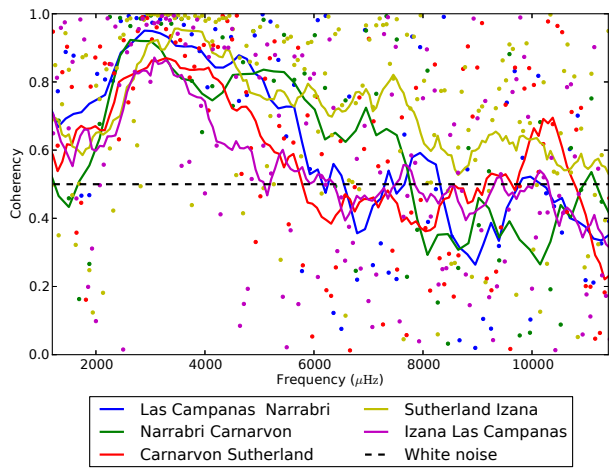}
	\caption{The coherency of calibrated contemporaneous data from BiSON stations adjacent in longitude for December 28 2007.  Dots show the unsmoothed coherency and solid lines show the coherency smoothed with a boxcar of width $1.2 \rm \, mHz$.  An idealised white noise coherency is show as a black horizontal dashed line.}
\label{fig::coh}
\end{figure}

We use the definition of coherence between two signals $x(t)$ and $y(t)$ as 
\begin{equation}
C_{xy} = \frac{|G_{xy}|^{2}}{G_{xx}G_{yy}},
\end{equation}
where $G_{xy}$ is the cross-spectral density and $G_{xx}$ and $G_{yy}$ are the autospectral densities.
Figure \ref{fig::golden} shows the excellent agreement between overlapping stations in the time domain.  Figure \ref{fig::coh} shows this agreement using the coherency of longitudinally adjacent stations with contemporaneous and continuous data.  The results show a high level of coherent signal around the $3 \rm \, mHz$ solar oscillations with near incoherent noise at very high frequency.\\
This means that stations as far away as different continents and different hemispheres measure the same solar oscillation signal at $3 \rm \, mHz$.  Yet the reduced coherency in the region that is dominated by solar granulation, $2 \rm \, mHz$ and below, demonstrates that each instrument does not observe that same solar background.  This allows us to combine contemporaneous data from different stations in the time domain and improve the signal-to-noise ratio of the modes of oscillation.  We combine coherent signal and nearly incoherent noise.  Clearly in the region of $3 \rm \, mHz$ with very high coherency and only a small contribution from the background, gains are marginal.  The true benefit comes when we consider regions in frequency where the background dominates.  Here we are reinforcing the tenuous signal and cancelling the dominant noise. \\  

\subsection{The full time series}
Having satisfied ourselves that the differential extinction correction is a sensible amendment to the residual calibration pipeline, and that corrected residuals have a high degree of coherency in the frequency regions of interest, we can consider the improvement in a long BiSON time series.  We pick a 22 year time series starting in 1991 when the network achieved maturity and duty cycles regularly achieved levels greater than $80 \%$.\\
\begin{figure}
\resizebox{\hsize}{!}{
\includegraphics{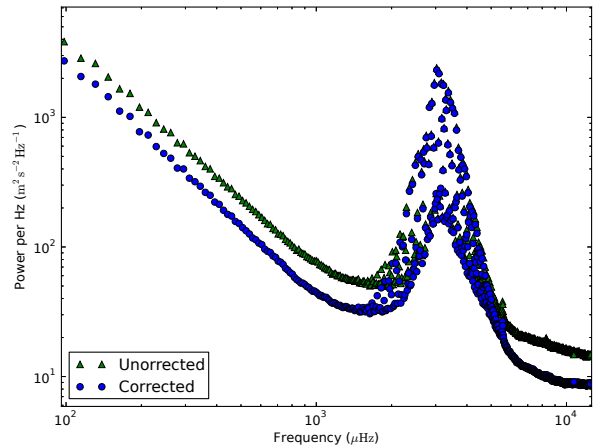}}
\caption{The BiSON 22 year power spectra smoothed over $16 \, \rm \mu Hz$.  Green dots show the power spectrum created with the old calibration procedure.  Blue dots show the new power spectrum created with a calibration including a correction for differential extinction and a the weighted averaging of contemporaneous stations.}
\label{fig::full_ps}
\end{figure}
Figure \ref{fig::full_ps} shows the smoothed frequency-power spectra for a 22 year time series, with and without the correction for differential extinction and weighted averaging of contemporaneous residuals.  The improvement in noise characteristics is obvious with the greatest improvement occurring in the low-frequency regime.\\
We may summarise the very-low frequency noise characteristics by quoting the average power in the range $95-100 \rm \, \mu Hz$.  This range provides a region of frequency space thought to be largely void of instrumental artefacts and a range narrow enough not to be confused by the gradient of the background.  Following the central limit theorem, by smoothing over 100 bins we approach a normal distribution in power for which, over the quoted range, the mean amplitude is $0.24 \, \rm cm \, s^{-1}$ with a standard deviation of $0.02 \, \rm cm \, s^{-1}$.  Given the predictions from \cite{2010A&ARv..18..197A}, we can see the BiSON very-low-frequency regime now has statistics approaching the theoretical limit of g-mode detection.\\ 

\section{Conclusions}
\label{conclusions}
In this paper we have detailed the methods now used to calibrate and combine the BiSON Sun-as-a-star data set into a high-quality time series ready to do new science.  We have demonstrated a correction through calibration for the effects of terrestrial atmospheric differential extinction and we have shown that, in the absence of large effects due to differential extinction, we can significantly improve the noise characteristics by combining contemporaneous data from different stations.\\
The consequences of the improvements to the BiSON data preparation pipeline are important.  We have demonstrated improvements in the high-frequency noise levels, but the most significant improvements are witnessed at very-low frequency.  The 22 years of BiSON measurements now present noise characteristics of $0.24 \, \rm cm \, s^{-1}$ in the g-mode region of $95-100 \rm \; \mu Hz$, close to the theoretical detection limit.  Future work will bring the full weight of Bayesian statistical tests designed to assess the detection of g modes to bear on the improved BiSON frequency-power spectrum.\\
Finally we emphasise that this work optimises an existing data set.  The steps taken here provide an observational network, one still observing but yet dating back to the late 1970's, with a updated and improved data set.  The historic nature of BiSON makes it an ideal tool to seismically study the Sun-as-a-star activity cycle.  In these times of unexpected extended solar minima and the subdued rising phase of cycle 24, an improved BiSON data set is indeed a valuable tool.\\  

\section{Acknowledgements}
We would like to thank all those who are, or have been, associated with BiSON.  
BiSON is funded by the Science and Technology Facilities Council (STFC).
We acknowledge the Leverhulme Trust for funding the `Probing the Sun: inside and out' project upon which this research is based.  The research leading to these results has received funding from the European Community’s Seventh Framework Programme (FP7/2007-2013) under grant agreement no. 312495 (SOLARNET).

\bibliographystyle{mn2e_new}

 \bibliography{refs}

\begin{thebibliography}{28}
\expandafter\ifx\csname natexlab\endcsname\relax\def\natexlab#1{#1}\fi

\bibitem[{{Appourchaux}(2011)}]{2011arXiv1103.5352A}
{Appourchaux} T., 2011, ArXiv e-prints

\bibitem[{{Appourchaux} {et~al}\mbox{.}(2010){Appourchaux}, {Belkacem},
  {Broomhall}, {Chaplin}, {Gough}, {Houdek}, {Provost}, {Baudin}, {Boumier},
  {Elsworth}, {Garc{\'{\i}}a}, {Andersen}, {Finsterle}, {Fr{\"o}hlich},
  {Gabriel}, {Grec}, {Jim{\'e}nez}, {Kosovichev}, {Sekii}, {Toutain}, \&
  {Turck-Chi{\`e}ze}}]{2010A&ARv..18..197A}
{Appourchaux} T. {et~al.}, 2010, \aapr, 18, 197

\bibitem[{{Basu} {et~al}\mbox{.}(2009){Basu}, {Chaplin}, {Elsworth}, {New}, \&
  {Serenelli}}]{2009ApJ...699.1403B}
{Basu} S., {Chaplin} W.~J., {Elsworth} Y., {New} R., {Serenelli} A.~M., 2009,
  \apj, 699, 1403

\bibitem[{{Belmonte} {et~al}\mbox{.}(1988){Belmonte}, {Elsworth}, {Isaak},
  {New}, {Palle}, \& {Roca Cort{\'e}s}}]{1988ESASP.286..177B}
{Belmonte} J.~A., {Elsworth} Y.~P., {Isaak} G.~R., {New} R., {Palle} P.~L.,
  {Roca Cort{\'e}s} T., 1988, in ESA Special Publication, Vol. 286, Seismology
  of the Sun and Sun-Like Stars, {Rolfe} E.~J., ed., pp. 177--179

\bibitem[{{Brookes}, {Isaak} \& {van der Raay}(1978){Brookes}, {Isaak}, \& {van
  der Raay}}]{1978MNRAS.185....1B}
{Brookes} J.~R., {Isaak} G.~R., {van der Raay} H.~B., 1978, \mnras, 185, 1

\bibitem[{{Broomhall} {et~al}\mbox{.}(2009){Broomhall}, {Chaplin}, {Elsworth},
  \& {New}}]{2009MNRAS.397..793B}
{Broomhall} A.~M., {Chaplin} W.~J., {Elsworth} Y., {New} R., 2009, \mnras, 397,
  793

\bibitem[{{Chaplin} {et~al}\mbox{.}(1997){Chaplin}, {Elsworth}, {Howe},
  {Isaak}, {McLeod}, {Miller}, \& {New}}]{1997A&AS..125..195C}
{Chaplin} W.~J., {Elsworth} Y., {Howe} R., {Isaak} G.~R., {McLeod} C.~P.,
  {Miller} B.~A., {New} R., 1997, \aaps, 125, 195

\bibitem[{{Chaplin} {et~al}\mbox{.}(1996){Chaplin}, {Elsworth}, {Isaak},
  {Lines}, {McLeod}, {Miller}, \& {New}}]{1996MNRAS.282L..15C}
{Chaplin} W.~J., {Elsworth} Y., {Isaak} G.~R., {Lines} R., {McLeod} C.~P.,
  {Miller} B.~A., {New} R., 1996, \mnras, 282, L15

\bibitem[{{Chaplin} {et~al}\mbox{.}(2003){Chaplin}, {Elsworth}, {Isaak},
  {Marchenkov}, {Miller}, \& {New}}]{2003ESASP.517..247C}
{Chaplin} W.~J., {Elsworth} Y., {Isaak} G.~R., {Marchenkov} K.~I., {Miller}
  B.~A., {New} R., 2003, in ESA Special Publication, Vol. 517, GONG+ 2002.
  Local and Global Helioseismology: the Present and Future, {Sawaya-Lacoste}
  H., ed., pp. 247--250

\bibitem[{{Cumberbatch} {et~al}\mbox{.}(2010){Cumberbatch}, {Guzik}, {Silk},
  {Watson}, \& {West}}]{2010PhRvD..82j3503C}
{Cumberbatch} D.~T., {Guzik} J.~A., {Silk} J., {Watson} L.~S., {West} S.~M.,
  2010, \prd, 82, 103503

\bibitem[{Davies(2011)}]{davies2011}
Davies G.~R., 2011, Investigating the low-frequency stability of bison's
  resonant scattering spectrometers

\bibitem[{{Davies} {et~al}\mbox{.}(2014){Davies}, {Broomhall}, {Chaplin},
  {Elsworth}, \& {Hale}}]{davies2014}
{Davies} G.~R., {Broomhall} A.~M., {Chaplin} W.~J., {Elsworth} Y., {Hale}
  S.~H., 2014, \mnras

\bibitem[{{de La Reza} \& {Mueller}(1975)}]{1975SoPh...43...15D}
{de La Reza} R., {Mueller} E.~A., 1975, \solphys, 43, 15

\bibitem[{{Egamberdiev} \& {Khamitov}(1991)}]{1991SoPh..133...81E}
{Egamberdiev} S., {Khamitov} I., 1991, \solphys, 133, 81

\bibitem[{{Elsworth} {et~al}\mbox{.}(1995){Elsworth}, {Howe}, {Isaak},
  {McLeod}, {Miller}, {New}, \& {Wheeler}}]{1995A&AS..113..379E}
{Elsworth} Y., {Howe} R., {Isaak} G.~R., {McLeod} C.~P., {Miller} B.~A., {New}
  R., {Wheeler} S.~J., 1995, \aaps, 113, 379

\bibitem[{{Garc{\'{\i}}a} {et~al}\mbox{.}(2007){Garc{\'{\i}}a},
  {Turck-Chi{\`e}ze}, {Jim{\'e}nez-Reyes}, {Ballot}, {Pall{\'e}},
  {Eff-Darwich}, {Mathur}, \& {Provost}}]{2007Sci...316.1591G}
{Garc{\'{\i}}a} R.~A., {Turck-Chi{\`e}ze} S., {Jim{\'e}nez-Reyes} S.~J.,
  {Ballot} J., {Pall{\'e}} P.~L., {Eff-Darwich} A., {Mathur} S., {Provost} J.,
  2007, Science, 316, 1591

\bibitem[{{Grec} \& {Fossat}(1979)}]{1979A&A....77..351G}
{Grec} G., {Fossat} E., 1979, \aap, 77, 351

\bibitem[{{Kasten} \& {Young}(1989)}]{1989ApOpt..28.4735K}
{Kasten} F., {Young} A.~T., 1989, \ao, 28, 4735

\bibitem[{{Khatami} \& {Fossat}(1994)}]{1994ExA.....4..253K}
{Khatami} M., {Fossat} E., 1994, Experimental Astronomy, 4, 253

\bibitem[{{Kozhevnikova} \& {Makarova}(1961)}]{1961SvA.....5..399K}
{Kozhevnikova} N.~I., {Makarova} E.~A., 1961, \sovast, 5, 399

\bibitem[{{Lazrek} \& {Hill}(1993)}]{1993A&A...280..704L}
{Lazrek} M., {Hill} F., 1993, \aap, 280, 704

\bibitem[{{Lopes} \& {Silk}(2010)}]{2010ApJ...722L..95L}
{Lopes} I., {Silk} J., 2010, \apjl, 722, L95

\bibitem[{{McLeod} \& {Isaak}(1988)}]{1988ESASP.286..223M}
{McLeod} D.~B., {Isaak} G.~R., 1988, in ESA Special Publication, Vol. 286,
  Seismology of the Sun and Sun-Like Stars, {Rolfe} E.~J., ed., pp. 223--225

\bibitem[{{Nishikawa}(1990)}]{1990ApJS...74..315N}
{Nishikawa} J., 1990, \apjs, 74, 315

\bibitem[{{Severnyi}, {Kotov} \& {Tsap}(1980){Severnyi}, {Kotov}, \&
  {Tsap}}]{1980A&A....88..317S}
{Severnyi} A.~B., {Kotov} V.~A., {Tsap} T.~T., 1980, \aap, 88, 317

\bibitem[{{Siegel} \& {Roth}(2014)}]{2014arXiv1401.6888S}
{Siegel} D.~M., {Roth} M., 2014, ArXiv e-prints

\bibitem[{{Snodgrass} \& {Ulrich}(1990)}]{1990ApJ...351..309S}
{Snodgrass} H.~B., {Ulrich} R.~K., 1990, \apj, 351, 309

\bibitem[{{Turck-Chi{\`e}ze} {et~al}\mbox{.}(2012){Turck-Chi{\`e}ze},
  {Garc{\'{\i}}a}, {Lopes}, {Ballot}, {Couvidat}, {Mathur}, {Salabert}, \&
  {Silk}}]{2012ApJ...746L..12T}
{Turck-Chi{\`e}ze} S., {Garc{\'{\i}}a} R.~A., {Lopes} I., {Ballot} J.,
  {Couvidat} S., {Mathur} S., {Salabert} D., {Silk} J., 2012, \apjl, 746, L12

\end{thebibliography}

\section{appendix}
\label{appendix}
\subsubsection{Modelling the effects of differential extinction}
In order to measure radial velocity $v_{r}$, the RSS compares
the solar Fraunh\"ofer line as a function of wavelength
($I_{\odot}(\lambda)$) to known laboratory atomic
transitions.  In addition, the RSS makes Sun-as-a-star observations so
measured intensities ($I_{m}$) are integrated over the whole solar surface, $S$.  We can
state this as a simple model:
\begin{equation}
I_{m} = \int_{S} \int_{\lambda_{0} - \delta\lambda}^{\lambda_{0} +
\delta\lambda} I_{\odot}(v_{r},\lambda) \; d\lambda
\; dS,
\end{equation}
where $\lambda_{0}$ is the central wavelength of the scattering atomic 
transition and $\delta \lambda$ describes the transition line
width.\\  
To achieve a more complete description of the system we must include 
atmospheric extinction suffered when making ground-based
observations.  Because the Sun is a discrete source and extinction is
known to be differential across the solar surface, we must multiply
the line function with an atmospheric function.  By treating
the solar surface as a two dimensional disc, parametrised by the orthogonal coordinates $x$ and
$y$, we can include the atmospheric extinction $T$ as a function of
position on the solar disc, giving  
\begin{equation}
I_{m} = \int_{S} \int_{\lambda_{0} - \delta\lambda}^{\lambda_{0} +
\delta\lambda} I_{\odot}(v_{r},\lambda) T(x,y) \; d\lambda
\; dx \; dy.
\end{equation}
To complete the description we include a non-uniform instrumental
response $U$.
The scattering of light, required by the instrument to
measure intensity, will only occur for wavelengths within the limits
of the resonance scattering profile.  In addition, the instrument has
a non-homogeneous response to position on the solar disc, making $U$ a
function of wavelength and position $U(\lambda,x,y)$.  Including the
instrumental weighting in our simple model gives 
\begin{equation}
I_{m} = \int_{S} \int I_{\odot}(v_{r},\lambda) \; T(x,y) \;
U(\lambda,x,y)\; d\lambda 
\; dx \; dy.
\label{simple}
\end{equation}
This provides our framework to continue as we study each of the terms in
the integral separately.
\subsubsection{Instrumental weighting}
A BiSON RSS is designed to provide intensity measurements over a very narrow range in wavelength.  This allows the Doppler shift of the full-disc solar absorption line to be determined and calibrated into line-of-sight velocities.  A consequence of the instrument design is a non-uniform weighting over the solar disc.  Nowhere can this be seen more clearly than in the difference between inferred velocities from different detectors in the instrument.  BiSON instruments typically have two detectors that look into the vapour cell from opposite sides, that is sides orthogonal to the optical beam line and axis of solar rotation.  As a consequence each detector is preferentially weighted to the receding or approaching solar limb producing a velocity measurement that differs by a d.c. offset of tens of meters per second.  Hence, instrumental weighting is a function of both position on the solar disc and more obviously wavelength.\\    
Estimates of the instrumental weighting of an RSS have recently been revised.  \citet{2009MNRAS.397..793B} recognised the impact of ``instrumental Doppler imaging'', an optical depth effect in an RSS's resonance scattering vapour cell, and determined the instrumental weighting empirically.  However, these results assumed that photons detected by the instrument would be resonantly scattered only once. \cite{davies2011} revised these results by adopting a Monte Carlo approach that included the previously neglected effects from detections of photons undergoing multiple scattering in the RSS vapour cell.  Full details can be found in the reference.\\
\begin{figure}
\resizebox{\hsize}{!}{
\includegraphics{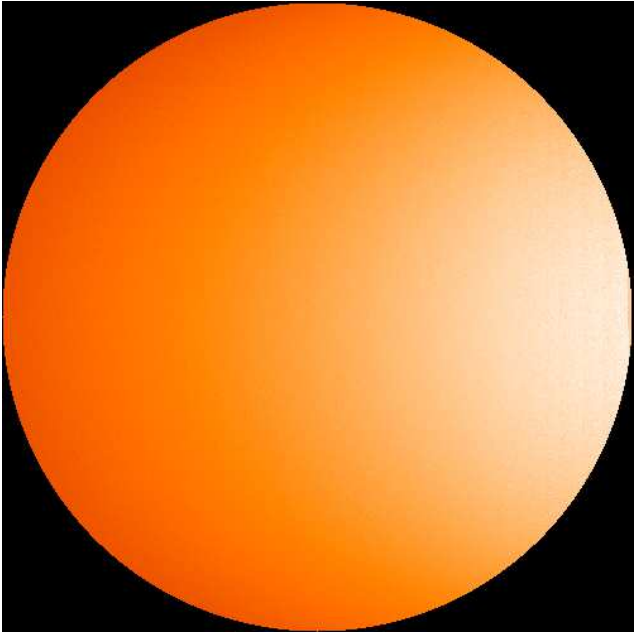}}
\caption{Simulated typical instrumental weighting to the solar disc for the starboard detector (on the right-hand side looking in on this diagram) for a BiSON resonance scattering spectrometer.}
\label{pos_inst_weight}
\end{figure}
The instrumental weighting function is constructed on a $500 \times 500$ pixel spatial map and results at a vapour temperature of 370 K are shown in figure \ref{pos_inst_weight}.\\

\subsubsection{Solar line function}
The opacity of the solar atmosphere places a spatial dependency in the solar line function due to sensitivity to varying depth, both in the terms of limb darkening and a variation in line parameters.
It is beyond the scope of this work to determine solar line parameters from a model of the solar atmosphere.  Instead, we take parameters from existing observations of the solar line and limb darkening.\\
For observational parameters we use those found by
\citet{1975SoPh...43...15D} using the McMath solar telescope at the
Kitt Peak Observatory.  These authors have taken centre-to-limb spectrogram observations at constant solar longitude (on the rotational axis) but
varying solar latitude, while avoiding the potential contamination of
active regions.  This work allows us to represent the solar line
observed at particular coordinates on the solar disc.  The spatial
integrations used $34800 \; \rm{km}$ in the y-direction and $174
\; \rm{km}$ in the x-direction, minimising the 
effects of rotational broadening, give a good representation of the
underlying solar line.  With only a very small observed asymmetry,
the solar line at any point on the solar disc can be well described by a
Gaussian absorption line with depth $d$, full width half maximum
(FWHM) $\Gamma$, and central wavelength $\lambda_{c}$.  The form of
each Gaussian line may be expressed algebraically as 
\begin{equation}
I_{\odot}(\lambda, \mu) = 1 - d(\mu) \exp\left[-4 \ln 2 \left(
\frac{(\lambda - \lambda_{c})^{2}}{\Gamma(\mu)^{2}}\right)
\right],
\label{iodot1}
\end{equation}
where $\mu$ describes the position on the solar disc such that
\begin{equation}
\mu = \cos\left(\sin^{-1}\left(\frac{\sqrt{x^2 +
y^2}}{R_{\odot}}\right) \right) = \sqrt{1 -
\left(\frac{\sqrt{x^{2}+y^2}}{R_{\odot}}\right)^{2}}. 
\label{mudef}
\end{equation}
The functions $d(\mu)$ and $\Gamma(\mu)$ can be crudely approximated across the solar disc (requiring both interpolation and extrapolation) by fitting a third order polynomial to the observed dependence of \citet{1975SoPh...43...15D}.\\  
To model the integrated solar line we construct a $500 \times 500$ pixel model of the solar disc by placing a Doppler shifted solar line function at each pixel.  Line-of-sight velocities are determined using the differential solar surface rotation function of \citet{1990ApJ...351..309S}.  
Because the introduction of a preferential axis (the rotational axis) breaks the symmetry of the solar surface, the alignment of the solar disc is now significant.  The orientation of the rotational axis with respect to heliographic coordinates can be described by two parameters.  The first is the solar position angle 
$P_{a}$, defined as the position angle between the geocentric north
pole and the solar rotational north pole measured eastward from
geocentric north.  The second, solar tilt angle $B_{0}$, is defined by
the heliographic latitude of the central point of the solar disc.  
The positions and orientations of solar system bodies are well
understood and $P_{a}$ and $B_{0}$ may be calculated or extracted
from the JPL ephemeris.\\  
Finally, the solar disc line function is completed with addition of the solar surface limb darkening function.  Results from \citet{1990ApJS...74..315N} give a limb darkening function at $770 \rm \; nm$ as 
\begin{equation}
\frac{I(\mu)}{I_{0}} = 0.469 + 0.7525\mu - 0.2215\mu^{2}.
\end{equation}
This limb darkening function is applied multiplicatively to the line function at each pixel.\\
\begin{figure}
\resizebox{\hsize}{!}{\includegraphics{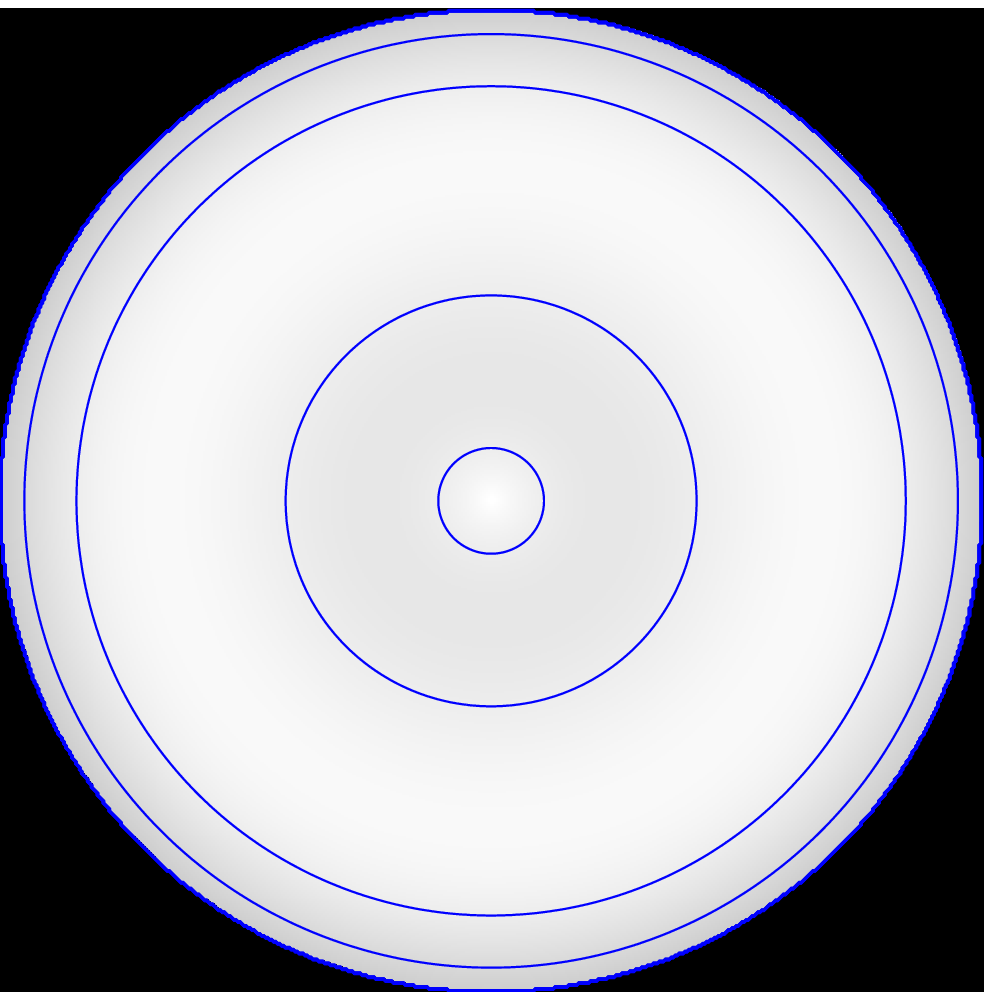}
\includegraphics{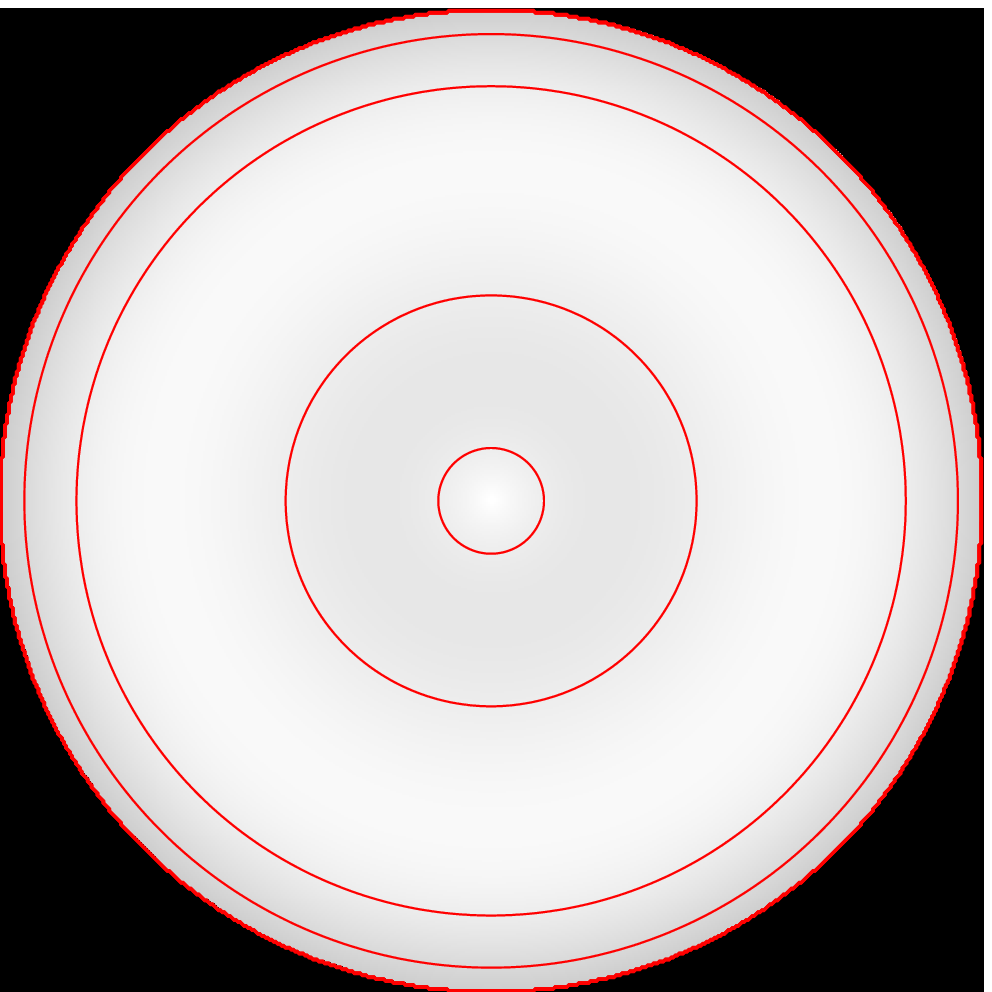}}
\resizebox{\hsize}{!}{\includegraphics{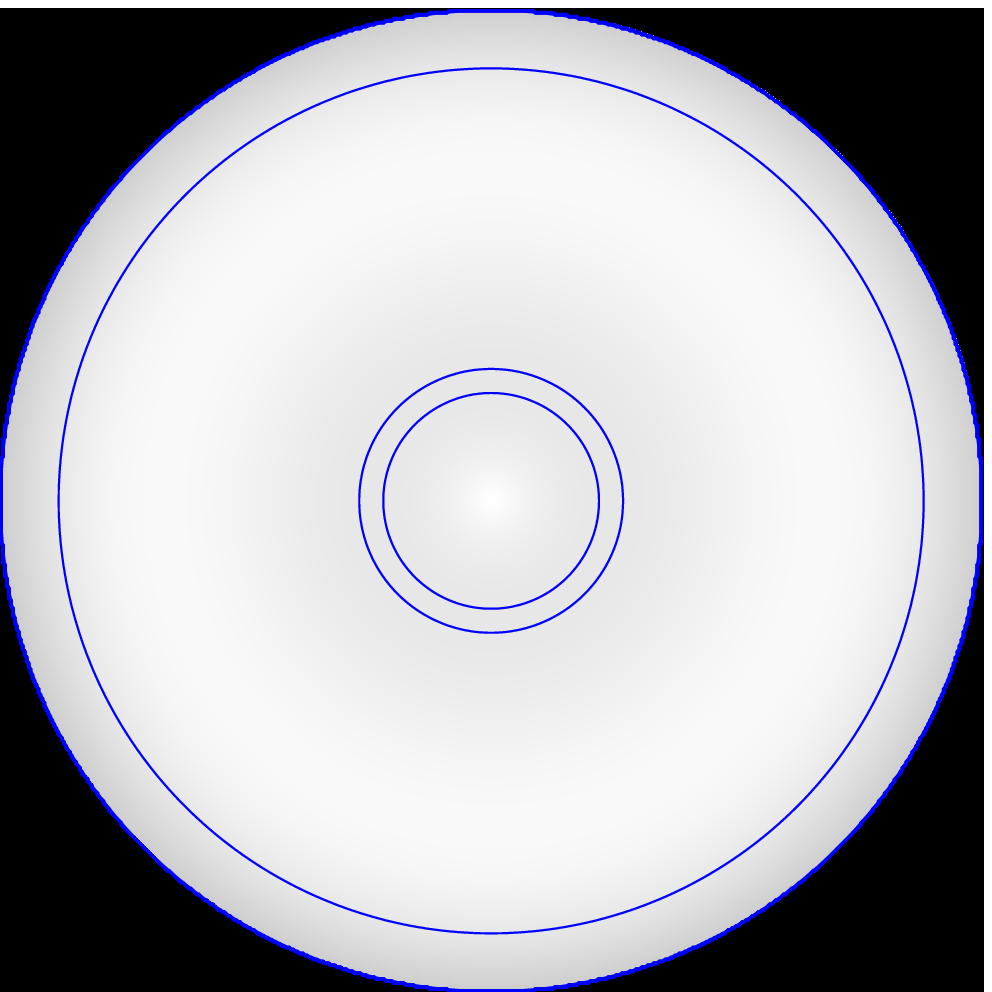}
\includegraphics{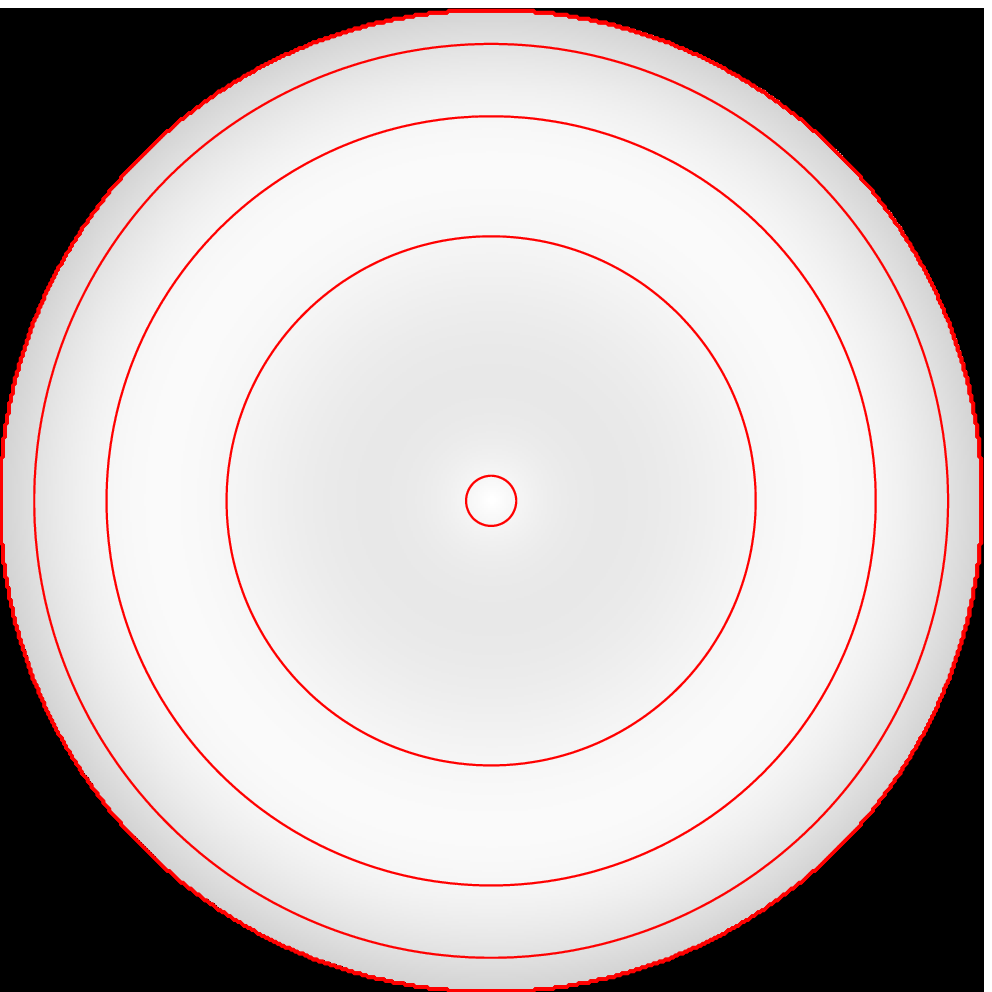}}
\resizebox{\hsize}{!}{\includegraphics{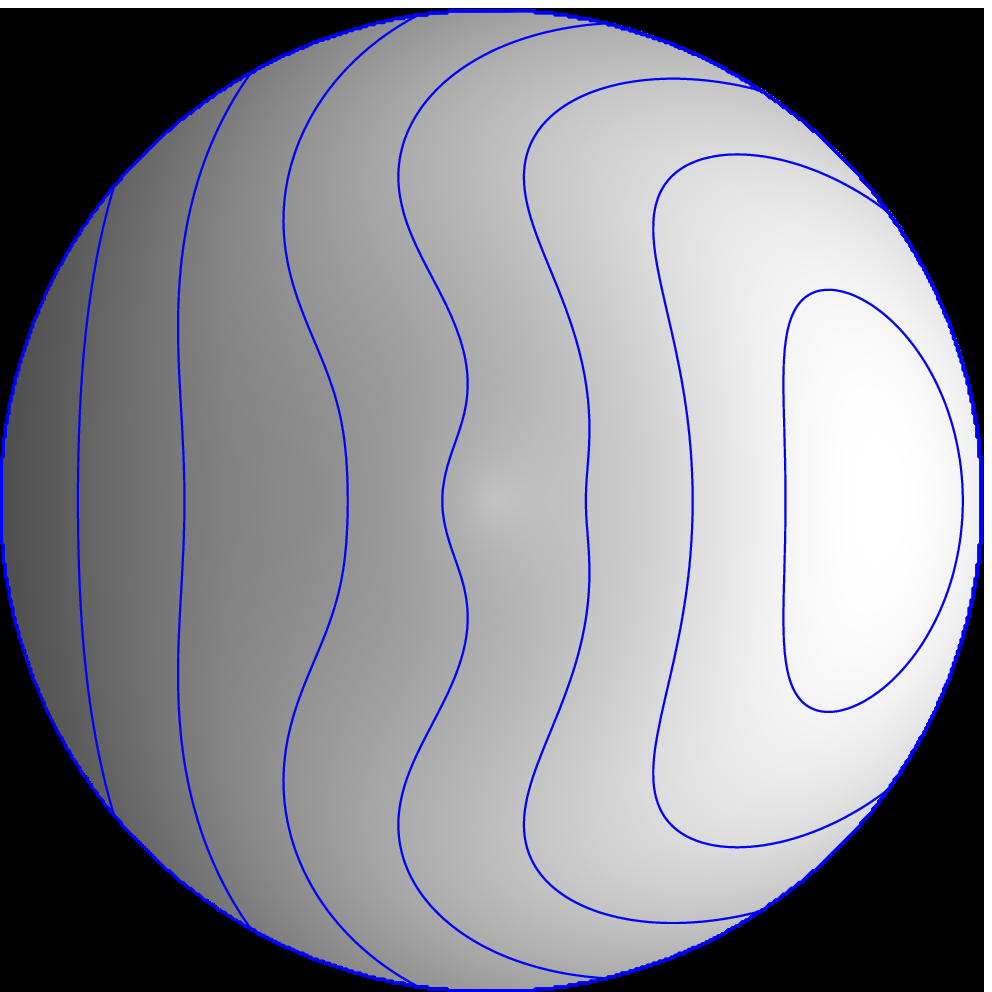}
\includegraphics{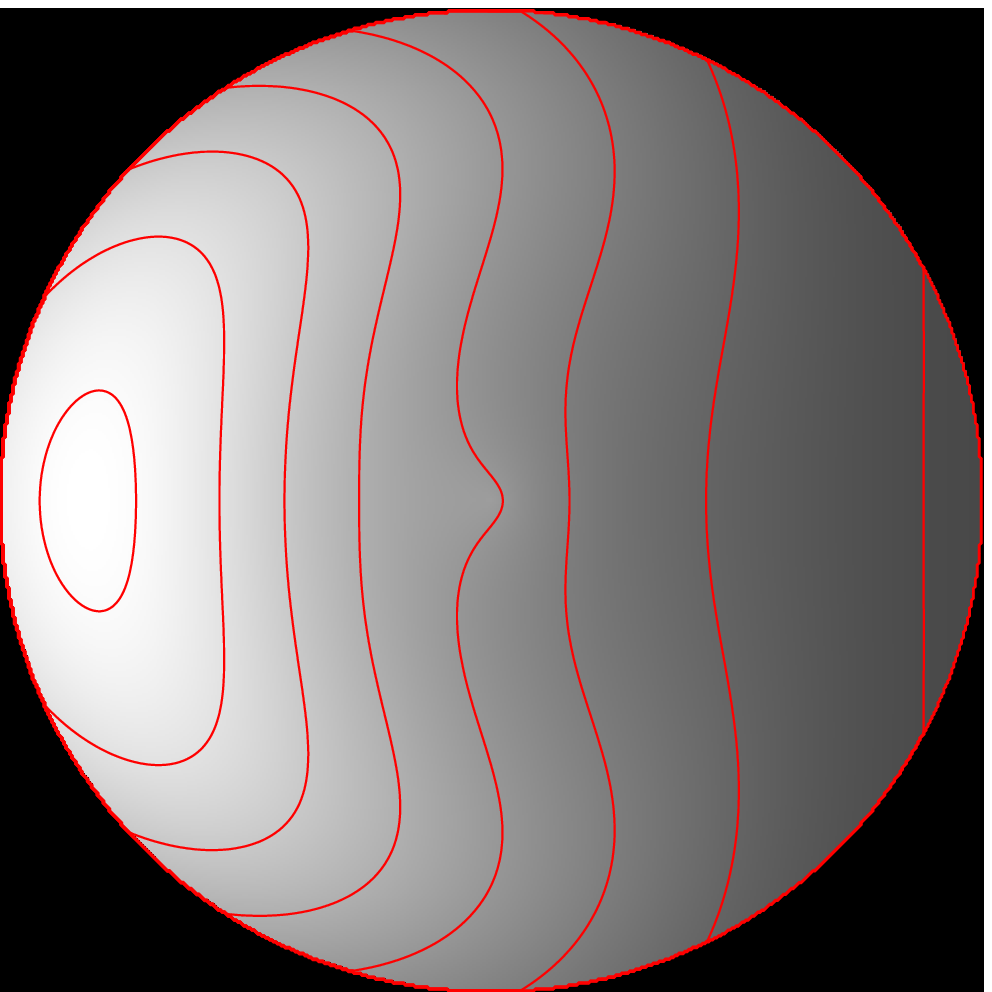}}
\resizebox{\hsize}{!}{\includegraphics{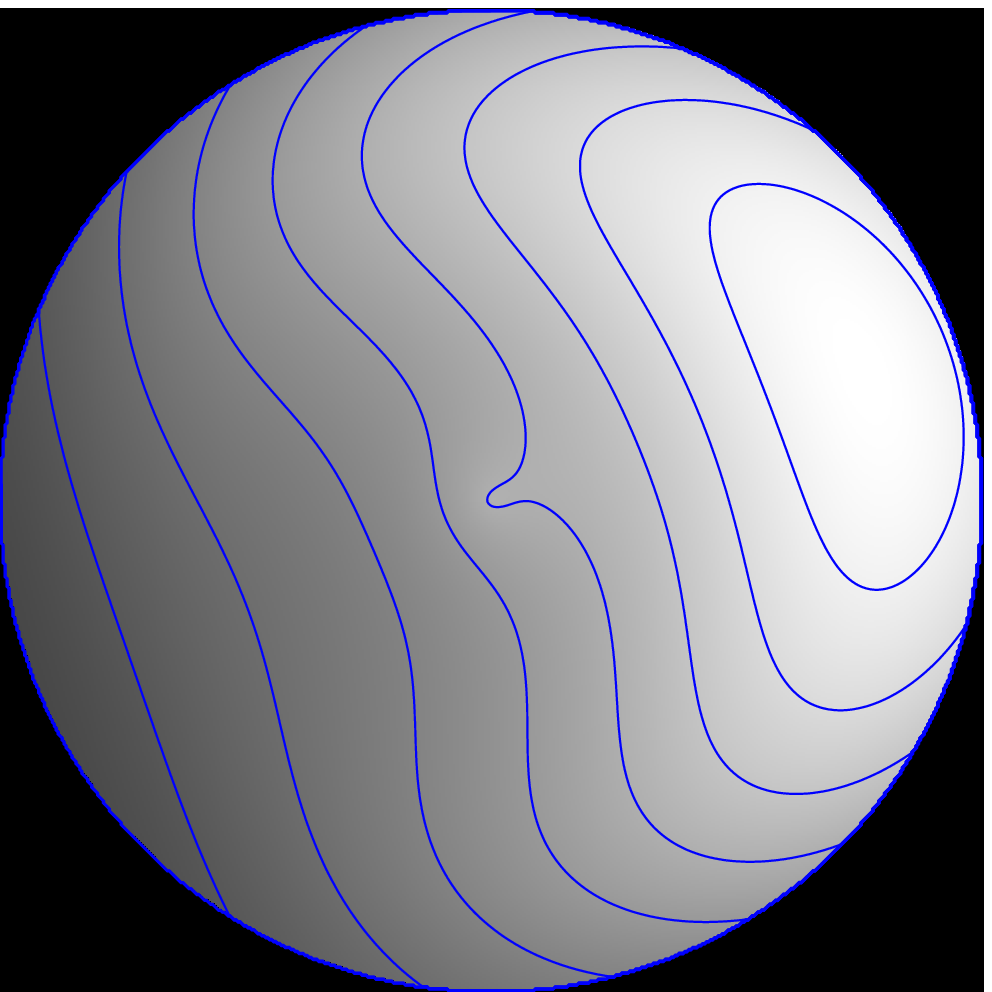}
\includegraphics{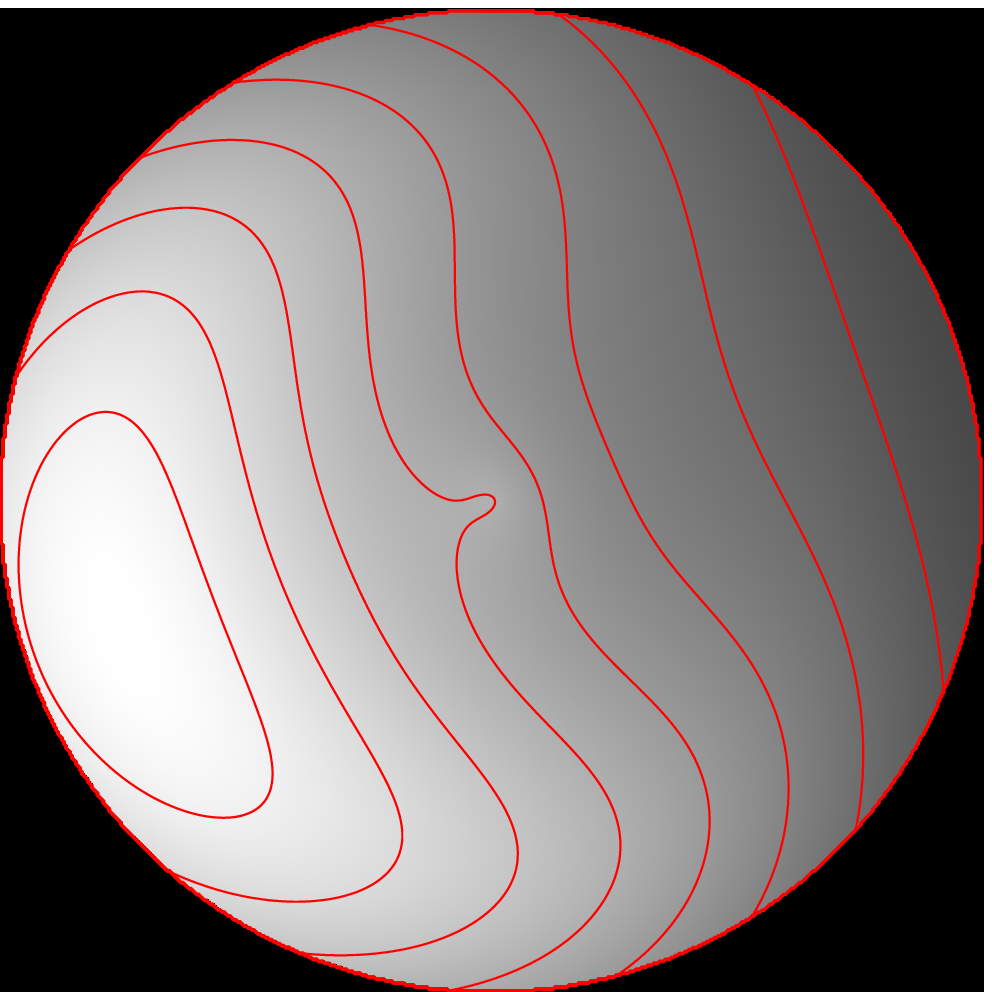}}
\resizebox{\hsize}{!}{\includegraphics{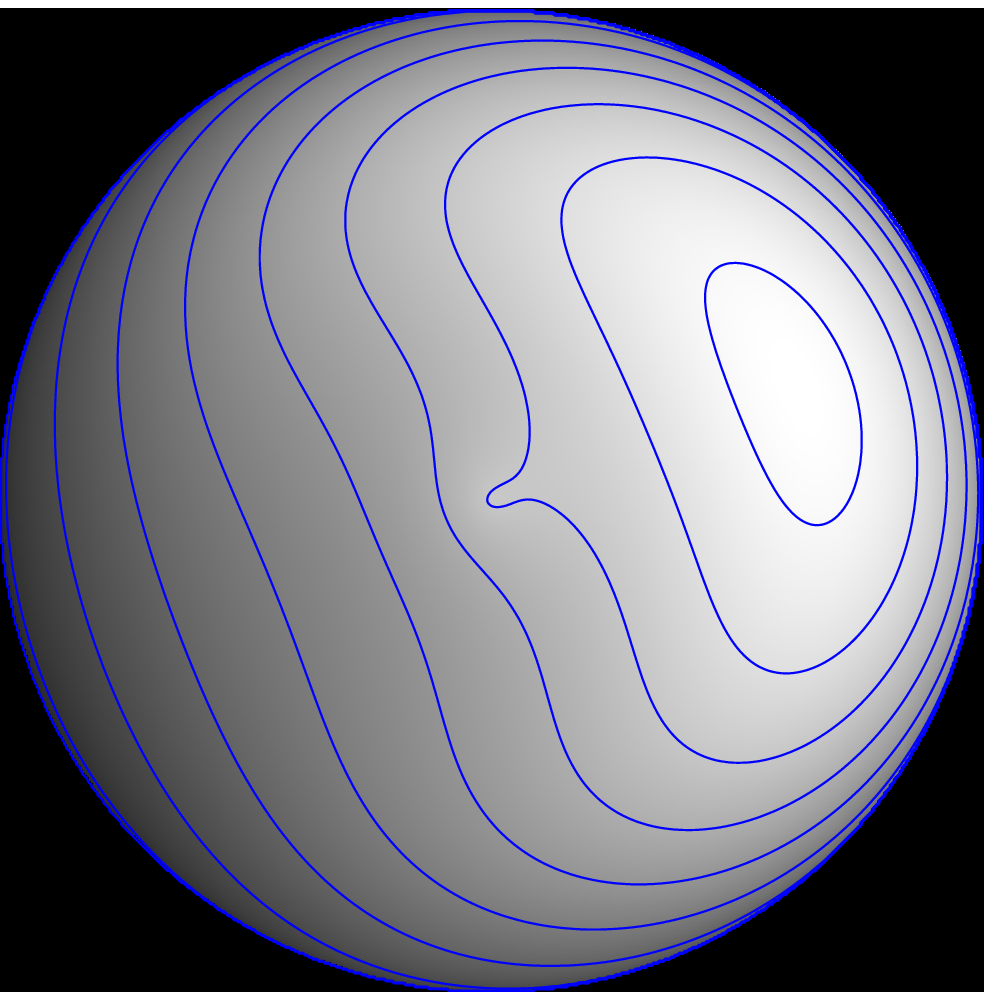}
\includegraphics{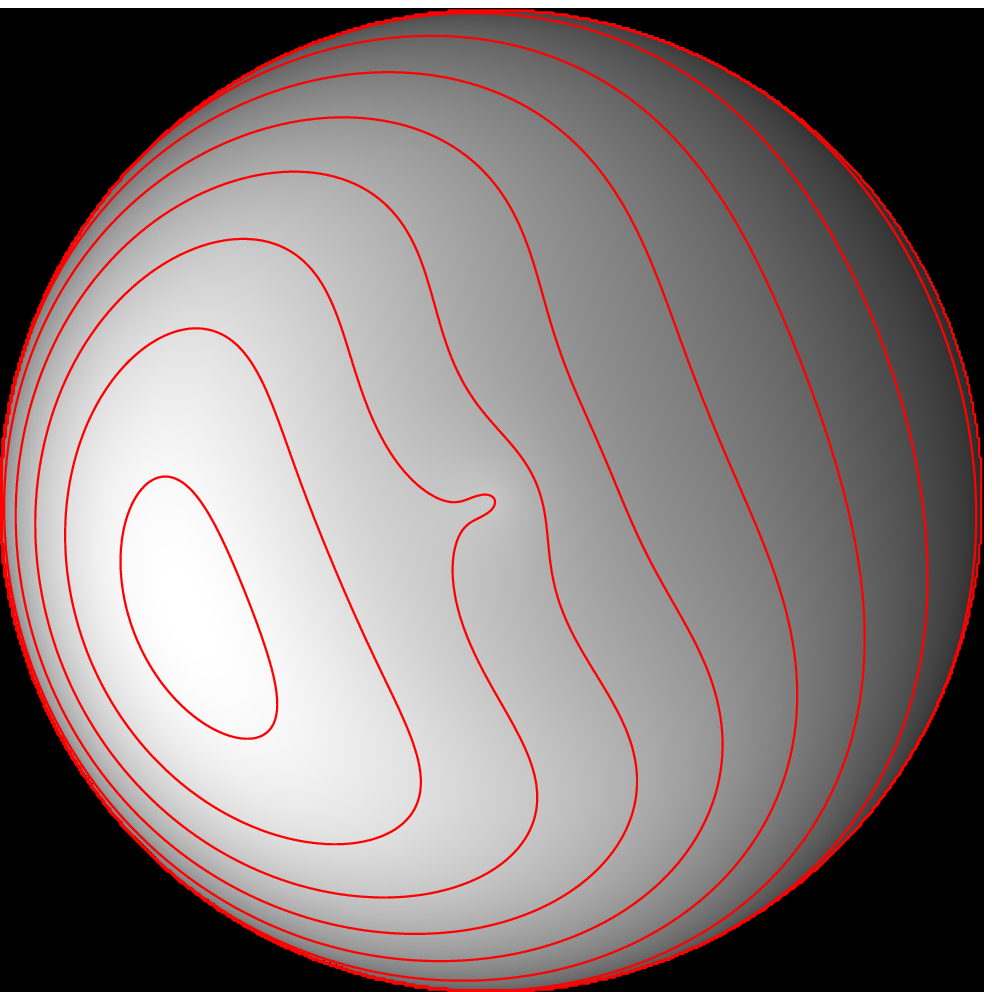}}
\caption{Simulated spatial weighting for the blue and red observational wings.  Top to bottom: base model; with line-of-sight velocity; with differential rotation; with position angle; and including limb darkening.}
\label{pos_line_weight}
\end{figure}

Figure \ref{pos_line_weight} shows results for the spatial weighting of the blue and red instrumental components, for the addition of each component of the model i.e. the underlying solar line, plus line-of-sight velocity, plus differential rotation, plus position angle ($P_{a} = 23.5^{\circ}$), and including limb darkening.  The detail seen at the center of the disc is a result of the changing solar line parameters in the de La Reza \& Mueller observations. 
\subsubsection{Atmospheric weighting to the solar disc}
Terrestrial atmospheric extinction can be modelled using the atmospheric extinction coefficient $\kappa$ and the airmass $A$ light must travel through to reach the observer.  The definition for the reduction in observed intensity ($I/I_{0}$) is one of exponential decay
\begin{equation}
\frac{I}{I_{0}} = \exp(-\kappa A).
\end{equation}
Normally the extinction coefficient would be defined as a function of wavelength, but the very narrow wavelength band over which RSS make observations allows us to treat $\kappa$ as a constant.\\
Airmass is defined as the path length that light from a point on the solar disc takes through the terrestrial atmosphere relative to the path length from a point source at the zenith.  A simple model for airmass is $A = \sec z_{0}$ where $z_{0}$ is the zenith angle.  Here we quote a better approximation in the form of an empirical relation found by \citet{1989ApOpt..28.4735K},
\begin{equation}
A = \frac{1}{\cos(z_{0}) + 0.50572(96.07995 - z_{0})^{-1.6364}}.
\end{equation}  
The solar zenith angle is formed between an observers local zenith and the line-of-sight to the centre point on the solar disc, and as such is clearly a function of time.  The solar zenith angle can be well approximated by either simple calculation or slightly more accurately extracted from the JPL ephemeris.\\
BiSON instruments do not view the Sun as a point source and hence the zenith angle varies across the solar disc, which is the source of differential extinction.  The zenith angle at any point on the disc can be calculated with knowledge of the solar zenith angle, the angular extent of the disc (taken as a fixed $0.5^{\circ}$), and the orientation of the disc with respect to the atmosphere.  Typically, BiSON instruments are mounted equatorially and this maintains a constant alignment of the instrument to the heliographic axes.  This allows us to consider the atmosphere rotating with respect to a fixed frame common to both the instrument and Solar disc.\\
The parallactic angle ($\eta$) describes the angle of the atmospheric gradient to the common Sun-instrument frame.  The parallactic angle is formally the angle between the great circle through a celestial object, the zenith, and the hour circle of the object, where the zenith is in the same direction as the atmospheric extinction gradient.  It is straightforward to show that the zenith angle of a point on the solar disc $z_{a}$, parametrised in $x$ and $y$, is
\begin{equation}
z_{a} = z_{0} + 0.25^{\circ}\frac{y \cos \eta - x \sin \eta}{R_{\odot}},
\end{equation}
where $R_{\odot}$ is the radius of the solar disc.\\
With this, we have sufficient information to calculate the atmospheric extinction at any point on the solar disc given the atmospheric extinction coefficient.  This can be applied as the final mask in the simple model defined earlier.  By evaluating our model, both with and without atmospheric extinction, we can determine the impact of differential extinction in the measured intensities.
\subsubsection{The differential extinction component}
In order to calibrate measured intensities into radial velocities, the blue ($I_B$) and red ($I_{r}$) components are combined in a ratio that maximises the sensitivity to velocity signals and minimises the sensitivity to atmospheric extinction.  This ratio $R$ is defined as 
\begin{equation}
R = \frac{I_b - I_r}{I_b + I_{r}}.
\end{equation}
\begin{figure}
\resizebox{\hsize}{!}{\includegraphics{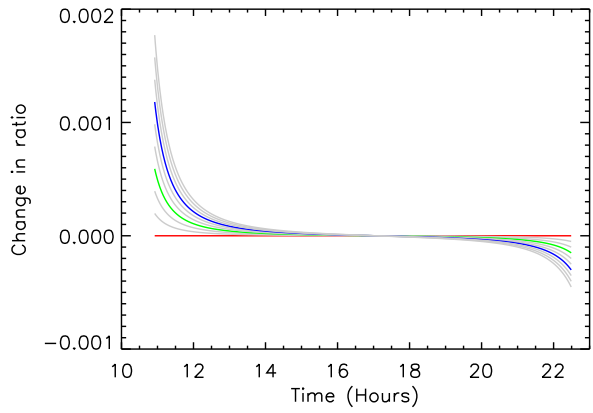}}
\caption{Differential extinction shape and magnitude for varying atmospheric absorption coefficient.  Highlighted: red $\kappa = 0$; green $\kappa = 0.03$; and blue $\kappa = 0.06$.}
\label{diff_ext}
\end{figure}
Figure \ref{diff_ext} shows the difference between the modelled ratios with and without atmospheric extinction for the BiSON Las Campanas station on 14 January 2009.  Hence we have uncovered the shape of the differential extinction component for various atmospheric extinction coefficients.\\
Further to this, we have generated 365 days of modelled data, including the appropriate Sun-observer line-of-sight velocity and other observer parameters, to give both the extinction affected daily ratios and the annual variation in differential extinction magnitude and shape.  Days are generated to match the duty cycle of the BiSON Las Campanas station through 2009.  We find that, to first order, the differential extinction shape is constant, but that the magnitude of the effect is dependent on the extinction coefficient and the line-of-sight velocity between the observer and Sun.  
Using this information we can propose a correction where shape is constrained but magnitude is a free parameter.

\label{lastpage}

\end{document}